\pdfoutput=1
\documentclass[%
 reprint,
superscriptaddress,
 amsmath,amssymb,
 aps,
 pra,
]{revtex4-2}

\usepackage{caption}
\usepackage{subcaption}
\usepackage{graphicx}
\usepackage{amsmath}
\usepackage{geometry}
\usepackage{hyperref}
\usepackage[T1]{fontenc}
\usepackage[utf8]{inputenc}
\usepackage{float}
\usepackage{csquotes}
\usepackage[version=4]{mhchem}
\usepackage{amssymb}
\usepackage{xcolor}
\usepackage{bbold}
\usepackage{dcolumn}
\usepackage{bm}
\usepackage{algorithm}
\usepackage{algpseudocode}

\begin{document}
\preprint{}

\title{
An Iterative Method to Improve the Precision of Quantum Phase Estimation Algorithm
}

\author{Junxu Li}
\email{Emial: lijunxu1996@gmail.com}
\affiliation{Department of Physics, College of Science, Northeastern University, Shenyang 110819, China}
\date{\today}

\begin{abstract}
Here we revisit the quantum phase estimation (QPE) algorithm, and devise an iterative method to improve the precision of QPE with propagators over a variety of time spans.
For a given propagator and a certain eigenstate as input, QPE with propagator is introduced to estimate the phase corresponding to an eigenenergy.
Due to the periodicity of the complex exponential, we can pinpoint the eigenenergy in a branch of comb-like ranges by applying QPE with propagators over longer time spans.
Thus, by picking up appropriate time spans, the iterative QPE with corresponding propagators can enable us to pinpoint the eigenenergy more precisely.
Moreover, even if there are only few qubits as ancilla qubits, high precision is still available by the proposed iterative method.
Our work provides a feasible and promising means toward precise estimations of eigenvalue on noisy intermediate-scale quantum (NISQ) devices. 
\end{abstract}

\maketitle
In 1982, Feynman firstly proposed to perform a computation with quantum mechanical devices\cite{feynman2018simulating}.
Since then, quantum computing has been of enormous interest, and is expected to solve the classically intractable problems, ranging from integer factorization to exploring many-body quantum physics.
Recently, landmarks in hardware developments since the `quantum supremacy' dawns\cite{arute2019quantum, zhong2020quantum, altman2021quantum, kim2023evidence, king2023quantum} considerably promote the trending to revisit and develop quantum algorithms\cite{biamonte2017quantum, bravyi2018quantum,  boixo2018characterizing, cong2019quantum, sajjan2022quantum}.

In this context, quantum phase estimation (QPE) algorithm,
firstly introduced by Kitaev in 1995\cite{kitaev1995quantum} to estimate the phase corresponding to an eigenvalue of a given unitary operator,
might be one of the most famous and powerful quantum algorithms.
QPE is also the critical subroutine in a variety of quantum algorithms, including Shor's algorithm\cite{shor1999polynomial} and  Harrow–Hassidim–Lloyd algorithm \cite{harrow2009quantum}.
Moreover, often combined with adiabatic state preparation\cite{farhi2000quantum}, QPE is also a powerful tool for chemists solving electronic structure problems, where QPE is implemented to estimate the ground energy of atoms and small molecules with given Hamiltonian\cite{aspuru2005simulated, lanyon2010towards,o2016scalable, cruz2020optimizing, bauer2020quantum, sajjan2022quantum}.
Even though, QPE requires deep circuits with ancilla qubits\cite{motta2020determining}, that are hard to execute in noisy intermediate-scale quantum (NISQ) era\cite{preskill2018quantum} with limited quantum resources.
For example, in 2019 IBM researchers implemented QPE on the IBM QX quantum computer, and reported that the accuracy and precision are severely constrained by the quantum devices' physical characteristics such as coherence time and error rates\cite{mohammadbagherpoor2019experimental}.
Therefore, a feasible approach to improve the precision of QPE is of great demand in NISQ era.

In this paper, we revisit the typical QPE, and devise an iterative method to improve the precision.
We will show that the QPE with propagators over longer time spans can lead to comb-like ranges in the estimation due to the periodicity of the complex phase.
Therefore, by implementing various propagators over appropriate time spans, the iterative QPE is capable to pinpoint the eigenenergy more precisely.
In the standard QPE algorithm, precision is gained by adding more qubits.
In the proposed iterative method, precision is gained in each iteration, even with only a few ancilla qubits.
Our work provides a feasible and promising means toward precise estimations of eigenvalue on NISQ devices. 
An outline for the rest of the paper is as follows.
In Sec.\ref{From propagators to unitary operators} we provide a brief overview of the implementation of propagators on quantum computer.
In Sec.\ref{Quantum Phase Estimation Algorithm with Propagators}  we revisit how QPE with propagators can be applied to solve the eigenvalue problem.
Then in Sec.\ref{An iterative method to improve the precision} we present the iterative method to improve the precision of QPE with propagators, and in Sec.\ref{Example} a simple example is presented.
Finally, we give our discussion and concluding remarks in Sec.\ref{Conclusions}.

\section{From propagators to unitary operators}
\label{From propagators to unitary operators}

For a given Hamiltonian $H$, the propagator describing the evolution from $t_{in}$ to $t_f$ can be written as,
\begin{equation}
    K(t_{f};t_{in})
    =
    \exp\left(iH(t_f-t_{in})\right)
\end{equation}
A general construction of the quantum circuit to simulate the propagator is performed as follows\cite{whitfield2011simulation}.
Initially, prepare $H$ as a sum over products of Pauli spin operators by the Jordan-Wigner transformation\cite{jordan1928pauli, somma2002simulating}.
Next, $H$ can be compiled into fundamental gates by using Trotter–Suzuki formulas\cite{hatano2005finding},
where the complicated Hamiltonian is complied as a sum of solvable ones.
Denote their sequential implementation on a quantum computer as $U_K(t_f; t_{in})$, which can be made to approximate the unitary propagator $K(t_f;t_{in})$.

The non-relativistic propagator can also be obtained in path integral formulation\cite{feynman1948space},
\begin{equation}
    K(x_f, t_f; x_{in}, t_{in})
    =
    \int_{x_{in}}^{x_f}\exp(iS[x])\mathcal{D}x
    \label{eq_propagator}
\end{equation}
where $S[x]$ denotes the action, and $\mathcal{D}x$ denotes the integration over all paths.

If the initial wavefunction $\Psi_{in}$ is known, the new wavefunction $\Psi_f$ can be derived with the propagator,
\begin{equation}
    \Psi_f(x_f, t_f)
    =
    \int K(x_f, t_f; x_{in}, t_{in})\Psi_{in}(x_{in}, t_{in}) dx_{in}
    \label{eq_psif}
\end{equation}
where the initial time $t_{in}$ and final time $t_f$ are fixed.
Recalling that the propagator can also be written as
\begin{equation}
    K(x_f, t_f; x_{in}, t_{in})
    =
    \sum_n |\psi_n\rangle\langle\psi_n|e^{-iE_n(t_f-t_{in})/\hbar}
\end{equation}
where $\psi_n$ is the $n-th$ eigenstate corresponding to eigenenergy $E_n$.
Thereafter the corresponding operator $U_K$ can be decomposed as,
\begin{equation}
    U_K(t_f;t_{in})=T^\dagger U_ET
\end{equation}
where $U_E=\sum_n|n\rangle\langle n| e^{iE_n(t_f-t_{in})/\hbar}$ is a diagonalized unitary operator, $T$ is a unitary transformation from the eigenstate basis to the computational basis.
In quantum simulation, the propagator can be efficiently carried out
using Trotter decomposition\cite{hatano2005finding}, provided that the Hamiltonian can be decomposed into a sum of local ones\cite{lloyd1996universal}.

\section{Quantum Phase Estimation Algorithm with Propagators}
\label{Quantum Phase Estimation Algorithm with Propagators}
The QPE with propagators is similar to the original QPE, where the given unitary operator $U$ is replaced by operator $U_K$ that approximates the propagator.

\begin{figure}[ht]
    \begin{center}
    \includegraphics[width=0.45\textwidth]{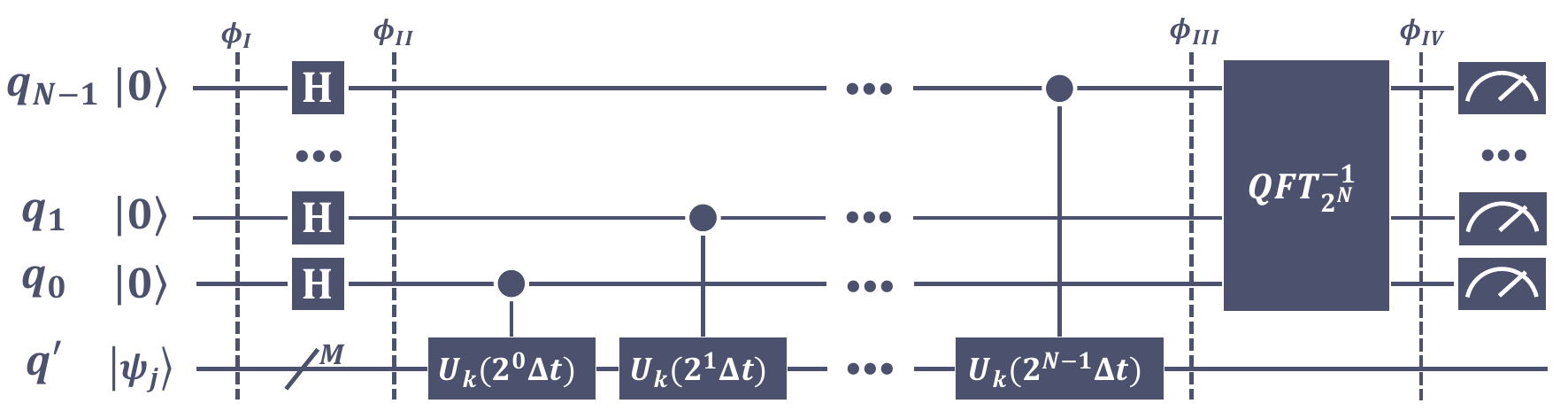}
    \end{center}
    \caption{
    {\bf Schematic diagram of the QPE circuit with propagators.}
    There are two registers, the first $N$ qubits initialized at ground state $|0\rangle$, and the others prepared at the $j-th$ eigenstate $|\psi_j\rangle$.
    At beginning, Hadamard gates are applied on $q_0,\cdot,q_{N-1}$ to convert the first register into a uniform superposition.
    The succeeding controlled gates then prepare a superposition of the evolution.
    Next, the inverse quantum Fourier transform is applied on the first register.
    By the end measurements are applied on the first register.
    We annotate the states at various stage as $\phi_I,\cdots, \phi_{IV}$, as indicated by the dashed line.
    }
    \label{fig_circuit}
\end{figure}

Our goal is to estimate the phase corresponding to an eigenvalue of the given propagator.
In Fig.(\ref{fig_circuit}) we present a schematic diagram of the QPE circuit with propagator.
There are two registers, the first $N$ qubits $q$ are initialized at ground state $|0\rangle$, and the others are prepared at $|\psi_j\rangle$, which is the $j-th$ eigenstate corresponding to eigenenergy $E_j$.
For simplicity, in the following discussion, we denote $U_K(\Delta t)=U_K(t_{in}+\Delta t; t_{in})$.

At the very beginning, the overall quantum state is
\begin{equation}
    |\Phi_I\rangle
    =
    |0\rangle^{\otimes N}\otimes|\psi_j\rangle
\end{equation}
Next, $N$ Hadamard gates prepare a uniform superposition, and the quantum state at stage II is
\begin{equation}
    |\Phi_{II}\rangle
    =
    \frac{1}{2^{N/2}}\sum_{n=0}^{2^N-1}|n\rangle\otimes|\psi_j\rangle
\end{equation}
The succeeding controlled gates prepare a superposition of the evolution, and the overall quantum state at stage III is
\begin{equation}
    |\Phi_{III}\rangle
    =
    \frac{1}{2^{N/2}}\sum_{n=0}^{2^N-1}|n\rangle\otimes U_K(n\Delta t)|\psi_j\rangle
\end{equation}
Recalling that $U_K(\Delta t)|\psi_j\rangle=e^{-iE_j\Delta t/\hbar}|\psi_j\rangle$, we have
\begin{equation}
    |\Phi_{III}\rangle
    =
    \frac{1}{2^{N/2}}\sum_{n=0}^{2^N-1}
    \left(
    |n\rangle\otimes e^{-iE_jn\Delta t/\hbar}|\psi_j\rangle
    \right)
\end{equation}
Then QFT is applied on the first $N$ qubits,
\begin{equation}
    \begin{split}
        |\Phi_{IV}\rangle
    =&
    \frac{1}{2^{N}}\sum_{x,n=0}^{2^N-1}
    \left(
    e^{2\pi inx/2^N}|x\rangle\otimes e^{-iE_jn\Delta t/\hbar}|\psi_j\rangle
    \right)
    \\=&
    \frac{1}{2^{N}}\sum_{x,n=0}^{2^N-1}
    \left(
    e^{2\pi inx/2^N-iE_jn\Delta t/\hbar}
    |x\rangle\otimes |\psi_j\rangle
    \right)
    \end{split}
\end{equation}
Notice that if $2\pi x/2^N = E_j\Delta t/\hbar$ exactly, 
\begin{equation}
    \frac{1}{2^{N}}\sum_{n=0}^{2^N-1}e^{2\pi inx/2^N-iE_jn\Delta t/\hbar}
    =
    1
\end{equation}
Otherwise we have
\begin{equation}
    \begin{split}
        &\frac{1}{2^{N}}\sum_{n=0}^{2^N-1}e^{2\pi inx/2^N-iE_jn\Delta t/\hbar}
    \\=&
    \frac{1}{2^{N}}
    \frac{1-e^{i(2\pi x-2^NE_j\Delta t/\hbar)}}{1-e^{i(2\pi x/2^N-E_j\Delta t/\hbar)}}
    \end{split}
\end{equation}
In brief, if ${2^NE_j\Delta t}/{2\pi\hbar}$ is exactly an integer, the probability to get result $|x\rangle$ after measuring the first $N$ qubits is
\begin{equation}
    \begin{split}
        Pr(x)
    =\begin{cases}
    1, &x=\frac{2^NE_j\Delta t}{2\pi\hbar}
    \\
    0
    ,&x\neq\frac{2^NE_j\Delta t}{2\pi\hbar}
    \end{cases}
    \end{split}
    \label{eq_pr}
\end{equation}
On the other hand, if ${2^NE_j\Delta t}/{2\pi\hbar}$ is not an integer, then we have
\begin{equation}
    Pr(x)=
    \left|
    \frac{1}{2^{N}}
    \frac{1-e^{i(2\pi x-2^NE_j\Delta t/\hbar)}}{1-e^{i(2\pi x/2^N-E_j\Delta t/\hbar)}}
    \right|^2
\end{equation}

Therefore, the highest probability outcome is the integer value $y$ closest to the $E_j$ phase term. 
Integer $y$ ensures that $|y/2^N-E_j\Delta t/2\pi\hbar|\leq1/2^{N+1}$, and it is proven that $Pr(y)\geq 4/\pi^2$\cite{cleve1998quantum}.
Theoretically, by measuring all $q$ qubits after applying the QPE algorithm, we can pinpoint $E_j\Delta t/2\pi\hbar$ in a certain slot $[(y_j-1/2)/2^N, (y_j+1/2)/2^N]$, where $0\leq y_j<2^N-1$ is the nearest integer corresponding to the greatest probability $Pr(y_j)$.
The algorithm must be run many times to collect enough data to identify the correct peak in the sampled distribution.

\section{An iterative method to improve the precision}
\label{An iterative method to improve the precision}

The inclusion of propagators enables an iterative method to improve the precision of QPE.
In the following discussion, we assume that the QPE is always accurate as expected.
Yet due to the existence of noises and errors, the measurement results of QPE might not peak at a single slim slot.
Instead, it often occurs that there is a plateau that covers several slots. 
In this regard, we only need to replace the single slot with the range of the plateau, and the iterative method still works.

Consider the $j-th$ eigenstate $|\psi_j\rangle$ as input, by performing the QPE with propagator $U_K(\Delta t)$ we can pinpoint the phase term $E_j\Delta t/2\pi\hbar$ in a certain slot,
\begin{equation}
    \frac{E_j\Delta t}{2\pi\hbar} \in 
    \left[\frac{y_j-1/2}{2^N}, \frac{y_j+1/2}{2^N}\right]
    \label{eq_ori}
\end{equation}
where $0\leq y_j<2^N-1$ is the nearest integer. 
For simplicity, here we assume that $\Delta t\leq 2\pi\hbar/E_j$ and $E_j>0$, which ensures $E_j\Delta t/2\pi\hbar\in [0, 1]$.

\begin{figure}[t]
    \begin{center}
        \includegraphics[width=0.45\textwidth]{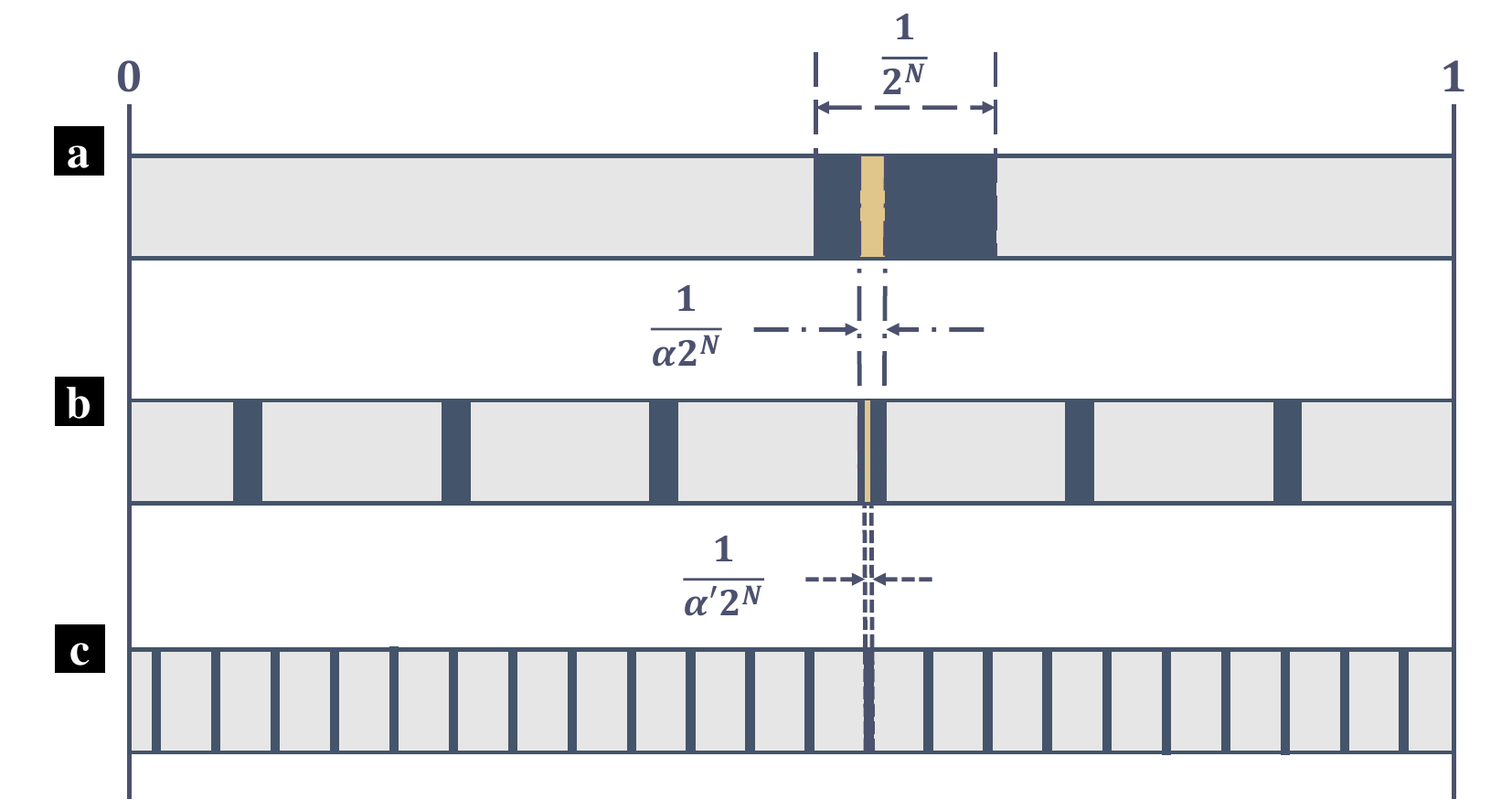}
    \end{center}
    \caption{
    {\bf Schematic diagram of the estimation $E_j\Delta t/2\pi\hbar$.}
    The estimation is colored in navy.
    (a) is the result of the QPE with propagator $U_K(\Delta t)$, where the estimation is located in a single stripe colored in navy, whose width is $1/2^N$.
    (b) is the result of QPE with propagator $U_K(\alpha\Delta t)$, whereas (c) for QPE with propagator $U_K(\alpha'\Delta t)$.
    In (b)(c) the estimated $E_j\Delta t/2\pi\hbar$ locates in a branch of narrower stripes.
    More precise estimation is achievable by studying the overlaps.
    Overlap of (a)(b) is colored in yellow, shown in (a).
    Overlap of (a)(b)(c) is colored in yellow, shown in (b).
    }
    \label{fig_slots}
\end{figure}

Similarly, if we apply the QPE with propagator over a longer time span $U_K(\alpha\Delta t)$, we can pinpoint $\alpha E_j\Delta t/2\pi\hbar$ in slot $[y_j(\alpha)/2^N, (y_j(\alpha)+1)/2^N]$, where $\alpha>1$, and $0\leq y_j(\alpha)<2^N-1$ is the corresponding nearest integer.
In Fig.(\ref{fig_slots}), we present a schematic diagram of the estimation. 
As shown in Fig.(\ref{fig_slots}a), the estimated ${E_j\Delta t}/{2\pi\hbar}$ is pinpointed in the stripe colored in navy, whose width is $1/2^N$.
Similarly, QPE with propagator $U_K(\alpha\Delta t)$ informs that
\begin{equation}
    \frac{\alpha}{2\pi} \left(\frac{E_j\Delta t}{\hbar} + 2k\pi\right) 
    \in \left[\frac{y_j(\alpha)-1/2}{2^N}, \frac{y_j(\alpha)+1/2}{2^N}\right]
    \label{eq_alpha}
\end{equation}
where $k\in Z$.
Due to the periodicity of the complex exponential, Eq.(\ref{eq_alpha}) can be rewritten as
\begin{equation}
    \frac{E_j\Delta t}{2\pi\hbar} \in 
    \left[\frac{y_j(\alpha)-1/2}{2^N\alpha}-\frac{k}{\alpha},
    \frac{y_j(\alpha)+1/2}{2^N\alpha}-\frac{k}{\alpha}\right]
    \label{eq_alpha1}
\end{equation}
Eq.(\ref{eq_alpha1}) indicates that the same estimation of $E_j\Delta t/2\pi\hbar$ locates in a branch of narrow stripes as shown in Fig.(\ref{fig_slots}b), whose widths are $1/\alpha 2^N$.
Notice that the exact value of $k$ can not be determined with Eq.(\ref{eq_alpha1}) itself. 
Recalling Eq.(\ref{eq_ori}), the estimated phase should also locate in single stripe with width $1/2^N$, as shown in Fig.(\ref{fig_slots}a).
Therefore, the $k$ values can only be the integers ensuring that the intersection of Eq.(\ref{eq_ori}) and Eq.(\ref{eq_alpha1}) is not empty.
Then we can pinpoint the phase $E_j\Delta t/2\pi\hbar$ more precisely, which is the overlap of Eq.(\ref{eq_ori}) and Eq.(\ref{eq_alpha1}).
The overlap is colored in yellow as depicted in Fig.(\ref{fig_slots}a).

Taking both Eq.(\ref{eq_ori}) and Eq.(\ref{eq_alpha1}) into account, we are able to make a more precise estimation than Eq.(\ref{eq_ori}) itself.
Yet we still need to find an optimal $\alpha$ value.
Intuitively, greater $\alpha$ value always leads to narrower stripes as shown in Eq.(\ref{eq_alpha1}).
However, greater $\alpha$ value also leads to closer neighbors (The distance between neighbors is $1/\alpha-1/2^N\alpha$, according to Eq.(\ref{eq_alpha1})).
If there are more than one narrow stripes in the overlap, we can hardly exclude anyone out of the final estimation.
Aiming for precise estimation, it should always be guaranteed that there is one and only one single stripe in the overlap.
Therefore, we have to make a trade-off, as greater $\alpha$ leads to narrower slots but closer neighbors.
Notice that if the distance between neighbors is no less than $1/2^N$, then there will be no more than one narrow stripe in the overlap.
In other words, we need to ensure that
\begin{equation}
    \frac{1}{\alpha}-\frac{1}{2^N\alpha}\geq \frac{1}{2^N}
\end{equation}
Thus the optimal choice is $\alpha= 2^N-1$, with which we can pinpoint the phase in a single stripe whose width is $1/2^N(2^N-1)$,
\begin{equation}
    \frac{E_j\Delta t}{2\pi\hbar} \in 
    \left[\frac{y'_j-1/2}{2^N(2^N-1)}, \frac{y'_j+1/2}{2^N(2^N-1)}\right]
    \label{eq_slot}
\end{equation}
where $y_j'\leq2^N(2^N-1)-1$ is the new nearest integer.

Moreover, higher precision is available by repeating the QPE with propagators and picking up appropriate $\alpha'$ values (Here we use notation $\alpha'$ to avoid confusions).
The QPE with propagator $U_K(\alpha'\Delta t)$ informs that
\begin{equation}
    \frac{E_j\Delta t}{2\pi\hbar} \in 
    \left[\frac{y_j(\alpha')-1/2}{2^N\alpha'}-\frac{k'}{\alpha'},
    \frac{y_j(\alpha')+1/2}{2^N\alpha'}-\frac{k'}{\alpha'}\right]
    \label{eq_alpha2}
\end{equation}
where $k'\in Z$, and $\alpha'>\alpha$.
As shown in Fig.(\ref{fig_slots}c), Eq.(\ref{eq_alpha2}) locate the estimation in much narrower stripes colored in navy, whose width is $1/\alpha'2^N$.
Similarly, the optimal choice of $\alpha'$ should be the maximum value ensuring that there is only one single stripe in the overlap among Eq.(\ref{eq_ori}), Eq.(\ref{eq_slot}) and Eq.(\ref{eq_alpha2}).
In other words, we need to keep the distance between new neighbors no less than the width of previous stripes, $1/2^N(2^N-1)$,
\begin{equation}
    \frac{1}{\alpha'}-\frac{1}{2^N\alpha'}\geq \frac{1}{2^N(2^N-1)}
\end{equation}
Then the optimal $\alpha'$ is obtained, $\alpha' = (2^N-1)^2$.
Recalling Eq.(\ref{eq_slot}), we can pinpoint $E_j\Delta t/2\pi\hbar$ in a narrower stripe colored in yellow as depicted in Fig.(\ref{fig_slots}b), whose width is $1/2^N(2^N-1)^2$, and
\begin{equation}
    \frac{E_j\Delta t}{2\pi\hbar} \in 
    \left[\frac{y''_j-1/2}{2^N(2^N-1)^2}, \frac{y''_j+1/2}{2^N(2^N-1)^2}\right]
    \label{eq_slot2}
\end{equation}
where $y_j''$ is the corresponding nearest integer.

We can always pick up appropriate $\alpha$ values, and the estimated ${E_j\Delta t}/{2\pi\hbar}$ will be pinpointed in narrower slots via QPE with propagators.
In other words, more precise estimation is available by repeating the iterative method.
In Tab.(\ref{tab}) we present the optimal $\alpha$ values for the iterative method and the corresponding width of the estimated ${E_j\Delta t}/{2\pi\hbar}$.
the $0-th$ iteration corresponds to the QPE with the original propagator $U_K(\Delta t)$, where $\alpha=1$.
The optimal $\alpha$ is the maximum value guaranteeing that there is only one single stripe in the overlap between the new estimation and the former ones.

\begin{table}
    \centering
    \begin{tabular}{c|c|c}
    \hline\hline
        Iterative Times  &Optimal $\alpha$   &Width \\
        \hline
         0&$1$  &$\frac{1}{2^N}$ \\
         1&$2^N-1$  &$\frac{1}{2^N(2^N-1)}$ \\
         2&$(2^N-1)^2$  &$\frac{1}{2^N(2^N-1)^2}$ \\
         &$\cdots$  &\\
         $n$&$(2^N-1)^n$  & $\frac{1}{2^N(2^N-1)^n}$\\
         \hline
    \end{tabular}
    \caption{\bf The optimal $\alpha$ values in iteration and the corresponding width of the corresponding estimation.}
    \label{tab}
\end{table}

In \ref{alg} we present the algorithm for the iterative method to improve the precision of QPEA, where $\epsilon_m$ is the maximum acceptable additive error, and the optimal $\alpha$ values are applied.

\begin{algorithm}[H]
\caption{Algorithm of the iterative QPEA}
\label{alg}
\begin{algorithmic}
\State {\bf Input:} Propagator $U_K(\Delta t)$, eigenstate $|\psi_j\rangle$,
\State \qquad maximum acceptable additive error $\epsilon_m$,
\State \qquad number of $q$ qubits $N$.
\State $\alpha \gets 1$
\State {\bf do} QPEA with $U_K(\alpha\Delta t)$, $|\psi_j\rangle$
\State {\bf do} Estimate the phase $E_j\Delta t/2\pi\hbar$,
\State \qquad calculate the additive error $\epsilon$.
\While{$\epsilon<\epsilon_m$}
    \State $\alpha \gets (2^N-1)\alpha$
    \State {\bf do} QPEA with $U_K(\alpha\Delta t)$, $|\psi_j\rangle$
    \State {\bf do} Estimate the phase $E_j\Delta t/2\pi\hbar$, 
    \State \qquad calculate the new additive error $\epsilon$.
\EndWhile
\State {\bf Return:} Phase $E_j\Delta t/2\pi\hbar$, additive error $\epsilon$.
\end{algorithmic}
\end{algorithm}

\section{Application to the two-site Hubbard Model}
\label{Example}
A simple example can be beneficial to showcase the strengths the proposed iterative method.
Consider the two-site Fermi Hubbard model, whose Hamiltonian is given by\cite{hubbard1963electron}
\begin{equation}
    H_{hub} = 
    -t\sum_\sigma{\left(c_{1,\sigma}^\dagger c_{2, \sigma}+c_{2, \sigma}^\dagger c_{1, \sigma}\right)}
    + u\sum_{j =1,2}n_{i,\uparrow}n_{i,\downarrow}
    \label{eq_hubbard}
\end{equation}
where $t$ denotes the transfer integral, $u$ denotes the on-site interaction, and $\sigma =\uparrow,\downarrow$ indicates the spin.
For simplicity, here we set $\hbar=1$, and $t=1$, $u=1$.
The eigenenergy, eigenstates and implementation of the propagator can be found in our recent work\cite{li2023toward}.

\begin{figure}[t]
    \begin{center}
        \includegraphics[width=0.45\textwidth]{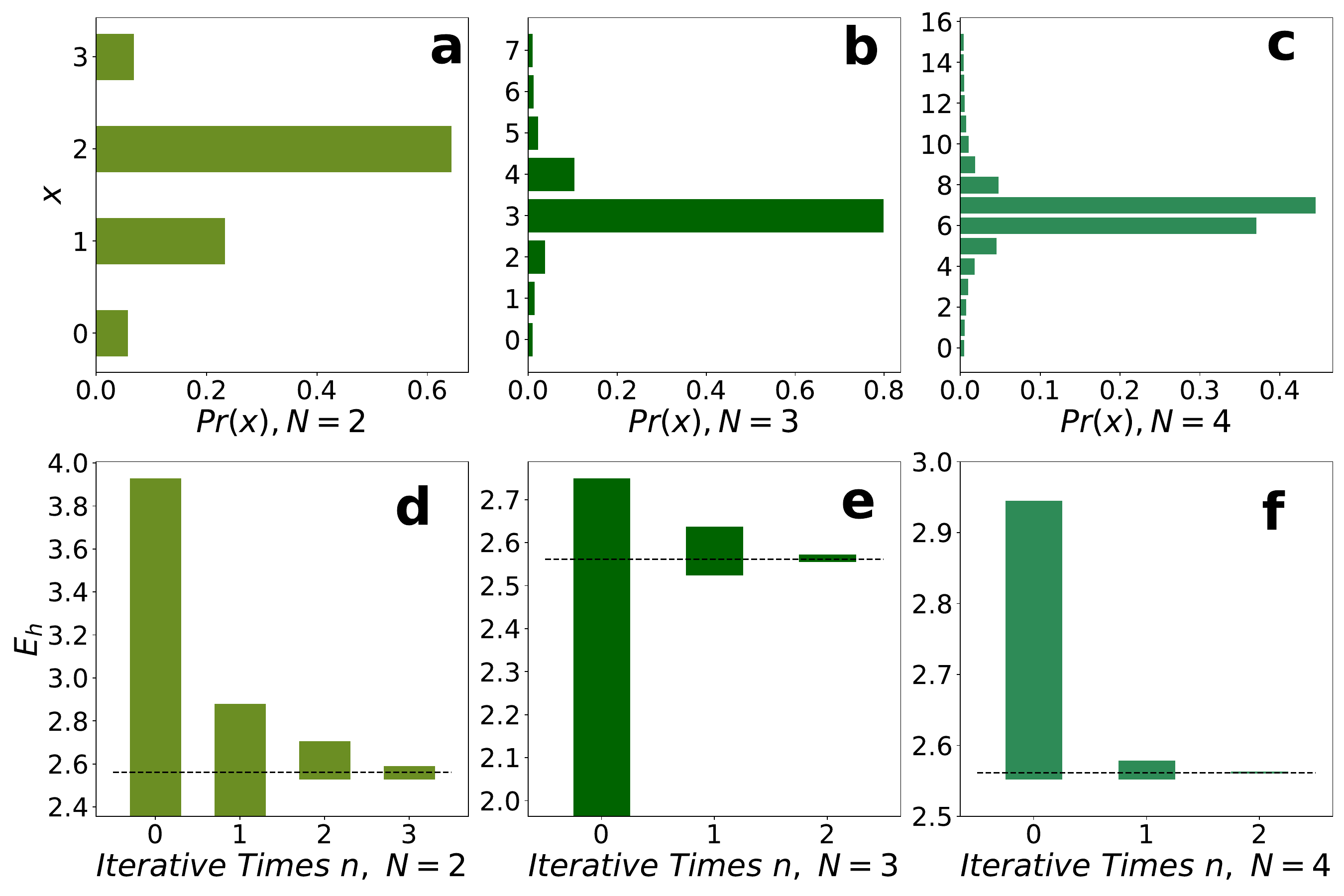}
    \end{center}
    \caption{
    {\bf Numerical simulation results for the standard QPE and the iterative method.}
    In (a, b, c) we present the numerical simulation of $Pr(x)$ for the original QPE, where $N=2,3,4$ is the number of qubits in the first register of QPE with propagators.
    In (d, e, f) we present the estimated eigenenergy $E_h$ with the iterative method, where $n$ is the iterative times, and $n=0$ corresponds to the original QPE.
    }
    \label{fig_simu}
\end{figure}

Here we focus on the highest energy state in half-filling model, corresponding to eigenenergy $E_h=2.56$.
In Fig.(\ref{fig_simu}a,b,c), we present the numerical simulation of $Pr(x)$ for the original QPE with $N=2,3,4$, where $Pr(x)$ indicate the probability to get result $x$ after measuring the first $N$ qubits, as shown in Eq.(\ref{eq_pr}).
The peaks correspond to the nearest integer, and we can pinpoint the phase term $E_h\Delta t/2\pi\hbar$ with Eq.(\ref{eq_ori}).
In Fig.(\ref{fig_simu}d, e, f) we present the estimated eigenenergy $E_h$ with the proposed iterative method, where $n$ is the iterative times, and $n=0$ corresponds to the standard QPE.
With Eq.(\ref{eq_slot}) and Eq.(\ref{eq_slot2}), we can pinpoint the eigenenergy $E_h$ in certain ranges as shown in the colored bars in Fig.(\ref{fig_simu}d, e, f).
The dashed line in Fig.(\ref{fig_simu}d, e, f) indicate the exact $E_h$.

In the standard QPE algorithm, precision is gained by adding more $q$ qubits.
Thus, greater $N$ leads to more precise estimation, corresponding to the $n=0$ bars in Fig.(\ref{fig_simu}d, e, f).
In the iterative QPE algorithm, precision is gained in each iteration.
Heights of the bars decrease exponentially with the iterative times $n$ in Fig.(\ref{fig_simu}d, e, f).
More $q$ qubits also leads to higher precision in the iterative QPE algorithm.
As shown in Fig.(\ref{fig_simu}d, e, f), heights of the bars decrease more quickly with greater $N$. 

\section{Discussions and Conclusions}
\label{Conclusions}

In general, precision of the final estimation is exponential to the iterative times and $N$, the number of qubits $q$.
Recalling Eq.(\ref{eq_slot}) and Eq.(\ref{eq_slot2}), the additive error of estimated ${E_j\Delta t}/{2\pi\hbar}$ under various iterative times $n$ and number of $q$ qubits $N$ can be given as
\begin{equation}
    \epsilon = \frac{C}{2^N(2^N-1)^n}
\end{equation}
where $C\geq 1$ is an integer, and in each iteration, we can pinpoint the peak of $Pr(x)$ in no more than $C$ slots. 
Ideally, there exists only one single peak slot in $Pr(x)$ and we can find a single nearest integer, where $C=1$.
However in real experiment, due to the existence of noises and imperfect hardware, it is often difficult to find out the exactly single peak slot of $Pr(x)$.
An example is as shown in Fig.(\ref{fig_simu}c), $Pr(x=7)$ is the peak in numerical simulation, yet $Pr(x=6)$ is very close to $Pr(x=7)$.
In experiment, due to noises and errors, one might find a plateau covering $Pr(x=6)$ and $Pr(x=7)$, and have to include both slots in the estimation.

\begin{figure}[t]
    \begin{center}
        \includegraphics[width=0.45\textwidth]{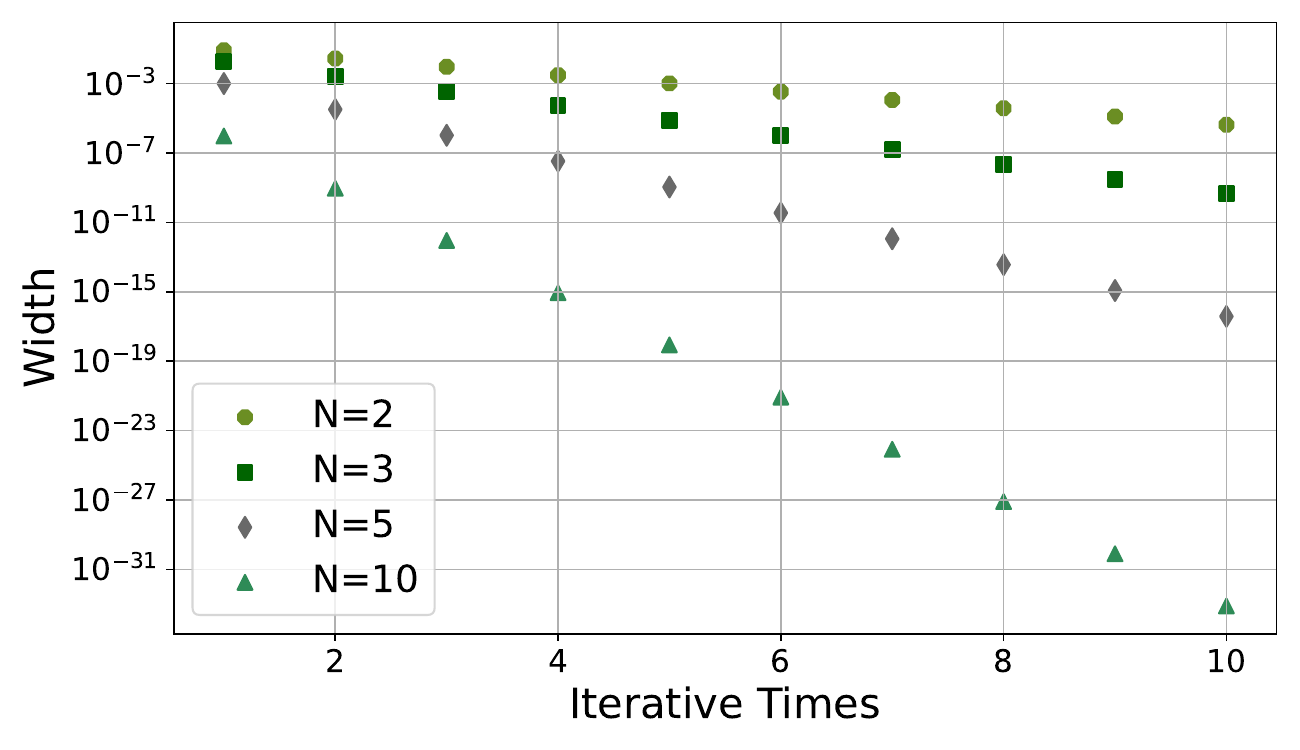}
    \end{center}
    \caption{
    {\bf Width of estimated $\frac{E_j\Delta t}{2\pi\hbar}$ under various iterative times and number of $q$ qubits.}
    $N$ is the number of qubits in the first register of QPE with propagators.
    Here we apply the optimal $\alpha$ in each iteration.
    }
    \label{fig_width}
\end{figure}

In Fig.(\ref{fig_width}), we present the width of estimated ${E_j\Delta t}/{2\pi\hbar}$ under various iterative times and number of $q$ qubits, where we always apply the optimal $\alpha$ in iteration.
We notice that high precision is available by the iterative method, even there are only 2 or 3 $q$ qubits as ancilla qubits, as shown in Fig.(\ref{fig_simu}d,e).
Thus, the iterative QPE with propagators is a promising approach toward precise estimations on NISQ devices, where it is often challenging to implement QPE among too many qubits for the existence of noises and limitation of connections.

Even though, the iterative method is hardly feasible for the original QPE with given operator $U$.
Here we focus on QPE with propagators, corresponding to a unitary operator $U_K(\Delta t)$.
In the iteration, for certain $\alpha$ value we do not directly implement $U_K(\Delta t)^\alpha$, yet we instead apply $U_K(\alpha\Delta t)$, which still, is a single unitary operator.
Therefore, the depth of quantum circuit does not change in the iteration.
However, if the given unitary operator $U$ in the original QPE is replaced as $U^\alpha$, the new quantum circuit can be considerably deeper, which can be extremely challenging to implement on NISQ devices.

In conclusion, we propose an iterative method to improve the precision of QPE.
We revisit the original QPE, and QPE with propagator enables us to pinpoint the phase corresponding to an eigenenergy in a certain range.
QPE with propagators over longer time spans can lead to `comb-like' estimations, with which we are able to pinpoint the same estimation in a branch of discrete narrower ranges.
Thus, by applying iterative QPE with propagators over a variety of time spans, we are able to pinpoint the estimation more precisely. 
High precision is available by the iterative method, even if there are only few qubits as ancilla qubits.
The iterative QPE with propagators provides a feasible and promising approach to estimate the corresponding eigenvalue more precisely on NISQ devices.

J.L gratefully acknowledges funding by National Natural Science Foundation of China (NSFC) under Grant No.12305012.

\bibliography{ref}

\end{document}